\documentclass[12pt]{article}
\usepackage{cite}
\usepackage{graphicx}
\usepackage{amssymb,amsfonts}
\usepackage{hangcaption}

\renewcommand{\theequation}
{\arabic{section}.\arabic{equation}}

\makeatletter
\def\eqnarray{ \stepcounter{equation} \let\@currentlabel=\theequation
 \global\@eqnswtrue
 \global\@eqcnt\z@
 \tabskip\@centering
 \let\\=\@eqncr
 $$\halign to \displaywidth\bgroup\@eqnsel\hskip\@centering
 $\displaystyle\tabskip\z@{##}$&\global\@eqcnt\@ne
 \hfil$\displaystyle{{}##{}}$\hfil
 &\global\@eqcnt\tw@$\displaystyle\tabskip\z@{##}$\hfil
 \tabskip\@centering&\llap{##}\tabskip\z@\cr}
\makeatother

\makeatletter
\def\@arrayacol{\edef\@preamble{\@preamble \hskip .5\arraycolsep}}
\def\array{\let\@acol\@arrayacol \let\@classz\@arrayclassz
\let\@classiv\@arrayclassiv \let\\\@arraycr\def\@halignto{}\@tabarray}
\makeatother
\makeatletter
\newcounter{subeqncnt}
\def\thesubeqncnt{\alph{subeqncnt}}
\def\subequations{\begingroup%
   \stepcounter{equation}\edef\@tempa{\theequation}%
   \let\c@equation\c@subeqncnt\c@subeqncnt\z@
   \edef\theequation{\@tempa\noexpand\thesubeqncnt}}

\makeatother
\newcommand{\captionfonts}{\small}
\makeatletter % Allow the use of @ in command names
\long\def\@makecaption#1#2{%
\vskip\abovecaptionskip
\sbox\@tempboxa{{\captionfonts #1: #2}}%
\ifdim \wd\@tempboxa >\hsize
{\captionfonts #1: #2\par}
\else
\hbox to\hsize{\hfil\box\@tempboxa\hfil}%
\fi
\vskip\belowcaptionskip}
\makeatother % Cancel the effect of \makeatletter
\newcommand{\del}{\partial}

\newcommand{\dd}{{\rm d}}

\def\imo{i}

\begin{document}

\setlength{\baselineskip}{7mm}
\begin{titlepage}
\begin{flushright}
{\tt APCTP-Pre2008-003} \\
{\tt arXiv:0808.2354[hep-th]} \\
September, 2008
\end{flushright}

\vspace{1cm}

\begin{center}
{\Large
Viscosity Bound, Causality Violation and Instability \\
with Stringy Correction and Charge}

\vspace{1cm}

{\sc{Xian-Hui Ge}}$^*$,
{\sc{Yoshinori Matsuo}}$^*$,
{\sc{Fu-Wen Shu}}$^*$\\
{\sc{Sang-Jin Sin}}$^{\dagger *}$ and
{\sc{Takuya Tsukioka}}$^*$

$*${\it{Asia Pacific Center for Theoretical Physics},} \\
{\it{Pohang, Gyeongbuk 790-784, Korea}} \\
{\sf{gexh, ymatsuo, fwshu,tsukioka@apctp.org}}
\\
$\dagger$ {\it{Department of Physics,}}
{\it{Hanyang University,}}
{\it{Seoul 133-791, Korea}} \\
{\sf{sjsin@hanyang.ac.kr}}
\end{center}

\vspace{1.5cm}

\begin{abstract}
Recently, it has been shown that if we consider the higher
derivative correction, the viscosity bound conjectured  to be
$\eta/s=1/4\pi$ is violated and so is the causality. In this paper,
we consider medium effect and the higher derivative correction
simultaneously by adding charge and Gauss-Bonnet terms. We find that
the viscosity bound violation is not changed by the charge. However,
we find that two effects together  create  another instability for
large momentum regime. We argue the presence of tachyonic modes and
show it numerically. The stability of the black brane requires the
Gauss-Bonnet coupling constant $\lambda$($=2\alpha'/l^2$) to be
smaller than $1/24$.
 We draw a phase diagram relevant to the
instability in charge-coupling space.
\end{abstract}

\end{titlepage}

%%%%%
\section{Introduction}
\setcounter{equation}{0}
\setcounter{footnote}{0}

After the discovery of consistency of AdS/CFT~\cite{ads/cft,gkp,w}
result and that of RHIC experiment on the viscosity/entropy-density
ratio~\cite{pss0,kss,bl},
much attention has been drawn to the calculational scheme provided
by string theory.
Some attempt has been made to map the entire process of RHIC
experiment in terms of the gravity dual~\cite{ssz}.
The way to include chemical potential in the theory was
figured out in~\cite{ksz,ht}.
Phases of these theories were also discussed
in D3/D7 setup \cite{nssy1,kmmmt,nssy2}
as well as in D4D8$\overline{\mbox{D8}}$\cite{ht}.

More recently, it had been conjectured that  the viscosity value of
theories with gravity dual  may  give a lower bound for the
$\eta/s=1/4\pi $ for all possible liquid\cite{kovtun}. However, the
authors of \cite{kp} and \cite{shenker} showed that if we consider
the stringy correction to $\alpha'$ order, the viscosity bound is
violated and causality is also \cite{shenker1} violated as a
consequence (See also for more recent paper \cite{neupane}).
% added %
%This inconsistency might come from the violation of AdS/CFT
%due to the stringy corrections which might require excited modes on
%strings.

The $\alpha'$ terms are also related to the (in)stability issues
of black holes.
The instability of $D$-dimensional asymptotically flat
Einstein-Gauss-Bonnet black holes has been discussed by
several authors\cite{dotti,konoplya}.
Their results show that for the gravitational perturbations
of Schwarzschild black holes in $D=(\mbox{from 5 to 11})$
Gauss-Bonnet gravity,
the instability occurs only for $D=5$ and $D=6$ cases
at large value of $\alpha'$ \cite{konoplya}.

In this paper, we add charge  together with the Gauss-Bonnet term, and
calculate $\eta/s$ and consider the stability issue  including the
causality violation.
%We find that the viscosity bound is still
%violated but the deviation is slightly reduced due to the charge effect.
We find that the viscosity bound violation is not changed by the
charge. However, we find that for large momenta regime, there exists
a new instability due to the charge effect. The linearized
perturbation has a negative frequency squared signaling an
instability. We draw the phase diagram  relevant to the instability.
The stability of the black brane requires $\lambda\leq 1/24$.
%The results infer that stability of the charged black branes
%requires $\eta/s\geq (5/6)(1/4\pi)$, which is closer to the
%conjectured viscosity bound $1/4\pi $ compared with $\eta/s\geq
% (16/25)(1/4\pi)$ obtained in \cite{shenker1}.
We emphasize that the new instability present only if both charge and   Gauss-Bonnet term  present.

The rest of the paper goes as follows.
In section 2, to set up we give a briefly review on the thermodynamic
properties of Reissner-Nordstr\"om-AdS black brane solution
in Gauss-Bonnet gravity.
In section 3, the Gauss-Bonnet correction to $\eta/s$ is calculated
via Kubo formula and its charge dependence is given in an explicit form.
In section 4, we study the causality violation problem for charged
black branes and reproduce the results found in Ref.\cite{shenker1}.
In section 5, we discuss the stability of Reissner-Nordstr\"om-AdS
black branes
in Gauss-Bonnet gravity.
Conclusions and discussions are presented in the last section.

\section{Reissner-Nordstr\"om-AdS black brane in Gauss-
\\Bonnet gravity}
\setcounter{equation}{0}
\setcounter{footnote}{0}

The thermodynamics and geometric properties of black objects in
Gauss-Bonnet gravity were studied in several papers
\cite{g1,g2,g3,cvetic,g4,ast}. In this section, we mainly review the
basic features of charged black holes in Gauss-Bonnet gravity.
Further details can be found in \cite{cvetic}.

We start by introducing the following action in $D$ dimensions
which includes Gauss-Bonnet terms and $U(1)$ gauge field:
\begin{equation}
\label{action}
I=\frac{1}{16 \pi G_{D}}\!\int\!\dd^{D}\!x
\sqrt{-g}\Big(R-2\Lambda+\alpha'\left(R_{\mu\nu\rho\sigma}
R^{\mu\nu\rho\sigma}-4R_{\mu\nu}R^{\mu\nu}+R^2\right)-4 \pi G_{D}
F_{\mu\nu}F^{\mu\nu}\Big),
\end{equation}
where $\alpha'$ is a (positive) Gauss-Bonnet coupling constant
with dimension $\rm(length)^2$ and the field strength is defied
as $F_{\mu\nu}(x)=\del_\mu A_\nu(x)-\del_\nu A_\mu (x)$.
The corresponding Einstein equation leads
\begin{eqnarray}
\label{einstein}
R_{\mu\nu}-\frac{1}{2}g_{\mu\nu}R+g_{\mu\nu}\Lambda
=8\pi G_{D}\Big(F_{\mu\rho}F_{\nu\sigma}g^{\rho\sigma}
-\frac{1}{4}g_{\mu\nu}F_{\rho\sigma}F^{\rho\sigma}\Big)
+T^{\rm eff}_{\mu\nu},
\end{eqnarray}
where
\begin{eqnarray}
T^{\rm eff}_{\mu\nu}
=\alpha'
\Bigg[
&&
\frac{1}{2}g_{\mu\nu}
\Big(R_{\alpha\beta\rho\sigma}R^{\alpha\beta\rho\sigma}
-4R_{\alpha\beta}R^{\alpha\beta}
+R^2\Big)
-2RR_{\mu\nu}+4R_{\mu\rho}R_\nu{}^{\rho}
\nonumber\\
&&
+4R_{\rho\sigma}R_{\mu\nu}{}^{\rho\sigma}
-2R_{\mu\rho\sigma\gamma}R_{\nu}{}^{\rho\sigma\gamma}\Bigg].
\end{eqnarray}
The charged black hole solution in $D$ dimensions for this action
is described by~\cite{cvetic}
\begin{subequations}
\begin{eqnarray}
\label{metric}
\dd s^2
&=&
\displaystyle
-H(r)N^2\dd t^2+H^{-1}(r)\dd r^2+\frac{r^2}{l^2}
h_{ij}\dd x^{i}\dd x^{j},
\\
A_t
&=&
\displaystyle
-\frac{Q}{4\pi(D-3)r^{D-3}},
\end{eqnarray}
\end{subequations}

\vspace*{-6mm}
\noindent
with
$$
H(r)=k_{0}+\frac{r^2}{2\alpha}\left(1-\sqrt{1-\frac{4\alpha}{l^2}\left(1-\frac{m
l^2 }{r^{D-1}}+\frac{q^2 l^2}{r^{2D-4}}\right)} \right), \quad
\Lambda =-\frac{(D-1)(D-2)}{2l^2},
$$
where $\alpha$ and $\alpha'$ are connected by a relation
$\alpha=(D-4)(D-3)\alpha'$ and the parameter $l$ corresponds to AdS
radius. The constant $N^2$ will be fixed later. Note that the
constant value of $k_0$ can be $\pm1$ or $0$ and $h_{ij}\dd x^{i}\dd
x^{j}$ represents the line element of a $(D-2)$-dimensional
hypersurface with constant curvature $(D-2)(D-3)k_{0}$ and volume
$V_{D-2}$. The gravitational mass $M$ and the charge $Q$ are
expressed as
\begin{eqnarray*}
M&=&\frac{(D-2)V_{D-2}}{16 \pi G_D  }m,
\\
Q^2&=&\frac{2\pi (D-2)(D-3)}{ G_D  }q^2.
\end{eqnarray*}
Taking the limit $\alpha'\rightarrow 0$ with $k_0=0$, the solution
may correspond to one for Reissner-Nordstr\"om-AdS (RN-AdS). The
hydrodynamic analysis in this background has been done
in\cite{gmsst}.

In the following, we mainly focus on five-dimensional case with
$k_0=0$. Defining $\lambda=\alpha/l^2(=2\alpha'/l^2)$, the function
$H(r)$ becomes
\begin{equation}
H(r)=\frac{r^2}{2\lambda l^2}\left[1-\sqrt{1-4\lambda
\left(1-\frac{r^2_{+}}{r^2}\right)
\left(1-\frac{r^2_{-}}{r^2}\right)
\left(1-\frac{r^2_{0}}{r^2}\right)}
\ \right],
\end{equation}
where $r_+$ and $r_-$ correspond to the outer and the inner horizons,
respectively, and $-r^2_0=r_+^2+r_-^2$.
The constant $N^2$ in the metric (\ref{metric}) can be fixed at the
boundary whose geometry would
reduce to flat Minkowski metric conformaly, i.e.\
$\dd s^2\propto -c^2\dd t^2+\dd\vec{x}^2$.
On the boundary $r\rightarrow\infty$, we have
$$
H(r)N^2 \rightarrow\frac{r^2}{l^2},
$$
so that $N^2$ is found to be
\begin{equation}
N^2=\frac{1}{2}\Big(1+\sqrt{1-4 \lambda}\ \Big).
\end{equation}
Note that the boundary speed of light is specified to
be unity $c=1$.

We shall give thermodynamic quantities of this background.
The temperature at the event horizon is defined as
\begin{equation}
T=\frac{1}{2\pi\sqrt{g_{rr}}}\frac{\dd \sqrt{g_{tt}}}{\dd
r}=\frac{Nr_{+}}{2\pi l^2}\left(2-\frac{q^2l^2}{r^6_{+}}\right).
\end{equation}
The black brane approaches extremal when
$q^2l^2/r^6_{+}\rightarrow2$ (i.e.\ $T\rightarrow 0$).
The entropy of RN-AdS black holes with Gauss-Bonnet terms can be
obtained by using $S=-\partial F/\partial T$,
where $F$ is the free energy.
After Wick rotation i.e.\ $t\rightarrow i \tau$,
the free energy can be obtained from the
action $I$ in (\ref{action}) through $F=-TI$.
By using an identity which is derived by taking trace over the
equation (\ref{einstein}),
$$
\alpha'\left(R^2-4R_{\mu\nu}R^{\mu\nu}
+R_{\mu\nu\rho\sigma}R^{\mu\nu\rho\sigma}\right)+3R-10\Lambda-4\pi G_5
g^{\mu\nu}g^{\rho\sigma}F_{\mu\rho}F_{\nu\sigma}=0,
$$
the action (\ref{action}) reduces to be on-shell,
\begin{eqnarray*}
I&=&
\frac{1}{16 \pi G_5}
\!\int^{\infty}_{r_{+}}\!\!\!\!\dd r
\!\int^{\frac{1}{T}}_{0}\!\!\!\!\dd\tau
\!\int\!\dd^3x
\sqrt{g}\Big(-2R+8 \Lambda\Big)
\nonumber
\\
&=&
\frac{1}{16 \pi G_5}\frac{V_3N}{Tl^3}
\!\int^{\infty}_{r_{+}}\!\!\!\!\dd r r^3
\left[\frac{2}{r^3}\left(r^3 H(r)\right)''-\frac{48}{l^2}\right]
\nonumber
\\
&=&
\frac{1}{16 \pi G_5}\frac{V_3N}{Tl^3}\left[ 2
\Big(r^3H(r)\Big)'\Bigg|^{\infty}_{r_{+}}
-\frac{12}{l^2}
r^4\Bigg|^{\infty}_{r_{+}}\right],
\end{eqnarray*}
where we used the scalar curvature $R=-\left(r^3H(r)\right)''/r^3$
%%%%% footnote %%%%%
\footnote{Note that when $k_0\neq 0$, the scalar curvature is
$R=-\left(r^3 H(r)\right)''/r^3+6k_0/r^2$.}.
%%%%%
%
Divergent terms arise in the action and we can regulate
the result by subtracting the action of the Gauss-Bonnet-modified
pure AdS space, which is obtained by setting $r_{\pm}=0$ and
$r_{0}=0$ in $H(r)$(no horizons), that is to say
\begin{eqnarray*}
I_{\rm AdS\mbox{-}GB}
&=&
\frac{1}{16 \pi G_5}
\!\int^{\infty}_{0}\!\!\!\!\dd r
\!\int^{\beta'}_{0}\!\!\!\!\dd\tau
\!\int\! \dd^3x
\sqrt{g}\Big(-2R+8 \Lambda\Big)
\nonumber
\\
&=&
\frac{1}{16 \pi G_5}\frac{\beta'V_3N}{l^3}
\left[
2\Big(r^3H(r)\Big)'\Bigg|^{\infty}_{0}
-\frac{12}{l^2}
r^4\Bigg|^{\infty}_{0}\right],
\end{eqnarray*}
where we assign a temperature $\beta'$ to AdS space with
Gauss-Bonnet terms which is \cite{w2,shenker}
$$
\beta'=\beta \left.
\left(\frac{g^{\rm BH}_{tt}}{g^{\rm AdS}_{tt}}\right)^{1/2}
\right|_{r=\infty}
=\frac{1}{T}
\left.
\left(\frac{g^{\rm BH}_{tt}}{g^{\rm AdS}_{tt}}\right)^{1/2}
\right|_{r=\infty}.
$$
Then we find
$$
\Delta I=I-I_{\rm AdS\mbox{-}GB}
=\frac{1}{16 \pi
G_5}\frac{12r^4_{+}V_3N}{Tl^5}-\frac{V_{3}Nr^3_{+}}{2 G_{5}l^3}.
$$
Finally, we obtain the entropy density\footnote{
%%%%% footnote %%%%%
For more general case, if $k\neq 0$, the Bekenstein-Hawking area law
is broken and the resulted entropy is given by
$\frac{V_{D-2}r^{D-2}_{+}}{4 G_{D} }\left(1+\frac{D-2}{D-4}\frac{2
\lambda l^2}{r^2_{+}}k_0\right)$ (see \cite{g3}).},
\begin{equation}
s=-\frac{1}{V_3}\frac{\partial F}{\partial
T}=\frac{1}{V_3}\frac{\partial (T\Delta I)}{\partial T}=\frac{1}{4
G_{5}}\frac{r^3_{+}}{l^3}.
\end{equation}

\section{Viscosity to entropy density ratio}
\setcounter{equation}{0}
\setcounter{footnote}{0}

Before considering a linearized perturbative theory in this
background, we shall summarize the basic procedure to calculate
Green function and the shear viscosity in Minkowski spacetime\cite{ss}.
We work on the five-dimensional background,
$$
\dd s=g_{m n}\dd x^m\dd x^n+g_{uu}\dd u^2,
$$
where $x^m=(t, x, y, z)$ and $u$ are the four-dimensional and the
radial coordinates, respectively. We refer the boundary as $u=0$ and
the horizon as $u=1$. A solution of the linearized equation of
motion may be given by,
$$
\phi(x, u)=\!\int\!\frac{\dd^4k}{(2\pi)^4}
\mbox{e}^{ikx}f_k(u)\phi^{(0)}(k),
$$
where the function $f_k(u)$ is normalized such that $f_k(0)=1$ at
the boundary.
An on-shell action might be reduced to surface terms in four dimensions
by using the equation of motion,
\begin{equation}
I[\phi^{(0)}]
=\!\int\!\frac{\dd^4k}{(2\pi)^4}\phi^{(0)}(-k){\cal G}(k,u)\phi^{(0)}(k)
\bigg|_{u=0}^{u=1},
\end{equation}
where the function ${\cal G}(k,u)$ can be written in terms of
$f_{\pm k}(u)$ and $\del_uf_{\pm k}(u)$.
Accommodating Gubser-Klebanov-Polyakov/Witten relation~\cite{gkp, w}
to Minkowski spacetime, Son and Starinets arrived at the
following formulation for
the retarded Green function:
\begin{equation}
G(k)=2{\cal G}(k, u)\bigg|_{u=0},
\label{green}
\end{equation}
where the incoming boundary condition at the horizon is imposed.
In this paper we consider the tensor type perturbation in the
background.
By using an obtained retarded Green function, one can estimate the
shear viscosity $\eta$ via Kubo formula,
\begin{equation}
\eta=-\lim_{\omega\rightarrow 0}
\frac{\mbox{Im}(G(\omega, 0))}{\omega}.
\label{kubo}
\end{equation}

Now let us proceed to calculate the shear viscosity by using Green
function in our background. As we see above, it is standard to
introduce new dimensionless coordinate $u=r^2_{+}/r^2$. The
five-dimensional metric with $k_0=0$ in (\ref{metric}) is then
deformed into
\begin{equation}
\dd s^2=\frac{-f(u)N^2\dd t^2+\dd{\vec{x}}^2}{l^2b^2u}
+\frac{l^2 \dd u^2}{4 u^2 f(u)},
\end{equation}
where
$$
f(u)=\frac{1}{2 \lambda}\left[1-\sqrt{1-4\lambda
(1-u)(1+u-au^2)}\ \right],
$$
and we denote $a\equiv q^2l^2/r^6_{+}$ and $b^2\equiv1/r^2_{+}$.
In this coordinate system,
the event horizon of the black brane is
at $u=1$, while $u=0$ is the boundary of the AdS space.

We now study small metric fluctuation
$h^{x}_{y}(t,z,u)\equiv\phi(t, z, u)$
around the black brane background of the form
\begin{equation}
\dd s^2
=\frac{-f(u)N^2\dd t^2+\dd{\vec{x}}^2+2\phi(t,z,u)\dd x\dd y}
{l^2b^2u}+\frac{l^2 \dd u^2}{4 u^2 f(u)}.
\end{equation}
By considering the spin under the $O(2)$ rotation in $(x, y)$-plane,
gauge perturbations would be decoupled within this tensor type
perturbation.
Using Fourier decomposition
$$
\phi(t, z, u)
=
\!\int\!\frac{\dd^4k}{(2\pi)^4}
\mbox{e}^{-i\omega t+ikz}\phi(k, u),
$$
we can obtain the following linearized equation of motion for
$\phi(u)$ from the equation (\ref{einstein}):
\begin{eqnarray}
0
&=&
\phi''(u)
+\frac{g'(u)}{g(u)}\phi'(u)
\nonumber
\\
&&
+\frac{\bar{\omega}^2}{uN^2f^2(u)}\phi(u)-\frac{\bar{k}^2\left[1-2\lambda
u^2\left(2u(u^{-1}f(u))''+3(u^{-1}f(u))'\right)\right]}{uf(u)\left[1+2\lambda
u^2(u^{-1}f(u))'\right]}\phi(u),
\label{maineq}
\end{eqnarray}
where
$$
g(u)= u^{-1}f(u)\left[1+2\lambda u^2 (u^{-1}f(u))'\right],
$$
$$
\bar{\omega}\equiv \frac{l^2b}{2}\omega,
\qquad
\bar{k}\equiv\frac{l^2b}{2}k,
$$
and the prime denotes the derivative with respect to $u$.

Let us solve the equation of motion (\ref{maineq}) in hydrodynamic
regime i.e.\ small $\omega$ and $k$.
We first impose a solution as
\begin{equation}
\phi(u)=(1-u)^\nu F(u),
\end{equation}
where $F(u)$ is a regular function at the horizon $u=1$,
so that the singularity at the horizon might be extracted.
Substituting this form into the equation of motion,
we can fix the parameter $\nu$ as $\nu=\pm i\omega/(4\pi T)$
where $T$ is the temperature.
We here choose
$$
\nu=-i\frac{\omega}{4\pi T},
$$
as the incoming wave condition.
In order to get the shear viscosity via Kubo formula (\ref{kubo})
by using Green function (\ref{green}),
it might be sufficient to consider series expansion of the solution
in terms of frequencies up to the linear order of $\omega(=i4\pi T\nu)$,
\begin{equation}
F(u)
=F_0(u)+\nu F_1(u) + {\cal O}(\nu^2, k^2).
\label{series}
\end{equation}
The equation of motion (\ref{maineq}) becomes the following
form up to ${\cal O}(\nu)$,
%%
%\begin{equation}
%\left[g(u)\left((1-u)^{\nu}F(u)\right)'\right]'=0,
%\end{equation}
%
%and up to ${\cal O}(\nu^2)$ we obtain
%
\begin{equation}
\label{eq}
\left[g(u)F'(u)\right]'
-\nu\left(\frac{1}{1-u}g(u)\right)'F(u)-\frac{2\nu}{1-u}g(u)F'(u)=0.
\end{equation}
Substituting the series expansion (\ref{series}) into the
equation (\ref{eq}), one can get the equations of motion for
$F_0(u)$ and $F_1(u)$ recursively.
From ${\cal O}(\nu^0)$ in the
equation (\ref{eq}), the equation for $F_0(u)$ is obtained as
\begin{equation}
\left[g(u)F'_0 (u)\right]'=0,
\end{equation}
and can be solved as
$$
F'_0 (u)=\frac{C_1}{g(u)},
$$
where $C_1$ is an integration constant.
Regularity of $F_0 (u)$ at the horizon implies that $C_1$ must be
zero as $g(u)$ goes to zero at the horizon.
Therefore, $F_0 (u)$ is a constant,
\begin{equation}
F_0(u)=C, \qquad (\mbox{const.}).
\end{equation}
The solution for $F_1 (u)$ can be obtained from equation (\ref{eq}) at
$\mathcal{O}(\nu^1)$,
\begin{equation}
\left[g(u)F'_1 (u)\right]'-\left(\frac{C}{1-u}g(u)\right)'=0.
\end{equation}
Integrating the above equation we get
\begin{equation}
F'_1 (u)=\frac{C}{1-u}+\frac{C_2}{g(u)}.
\end{equation}
The integration constant $C_2$ can be fixed by the regularity condition
of $F_1(u)$ at the horizon.
At the horizon $u=1$, the function $g(u)$ behaves as
$$
g(u)=\bigg((2-a)(1-2\lambda (2-a))\bigg)(1-u) +{\cal O}((1-u)^2).
$$
Then, the regularity condition at $u=1$ implies
\begin{equation}
C_2=-(2-a)(1-2\lambda (2-a))C.
\end{equation}
The remaining constant $C$ is estimated in terms of boundary value
of the field,
$$
\lim_{u\rightarrow 0}\phi(u)=\phi^{(0)},
$$
so that we could fix
\begin{equation}
C=\phi^{(0)}\Big(1+{\cal O}(\nu)\Big).
\end{equation}

\begin{figure}[bthp]
\begin{minipage}{1\hsize}
\vspace*{5mm}
\begin{center}
\includegraphics*[scale=0.5]{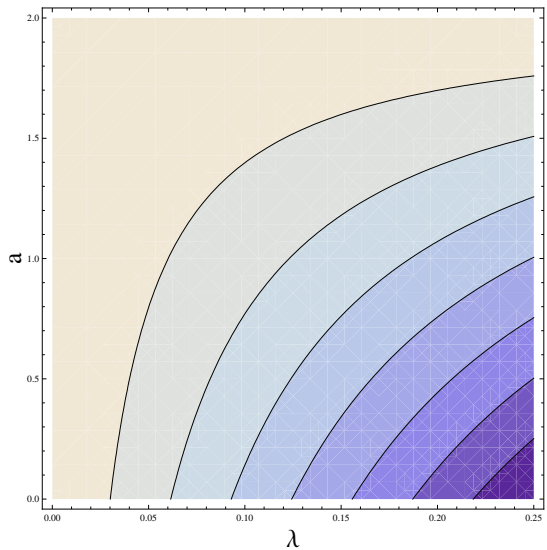}
\end{center}
\caption{Shear viscosity to entropy density ratio as a function of $a$
and $\lambda$. The lines correspond to $\eta/s=0.07, 0.06,...,0.01$,
respectively, from top to bottom. } \label{viscosity}
\end{minipage}
\end{figure}

Now we shall calculate the retarded Green function.
Using the equation of motion, the action reduces to the surface
terms.
The relevant part is given as
\begin{equation}
I[\phi(u)]=
-\frac{r^4_{+}N}{16\pi G_5 l^5}
\!\int\!\frac{\dd^4 k}{(2\pi)^4}
\Big(g(u)\phi(u)\phi'(u)+\cdots\Big)\Bigg|_{u=0}^{u=1}.
\end{equation}
Near the boundary $u=\varepsilon$,
using the obtained perturbative solution for $\phi(u)$,
we can get
\begin{eqnarray}
\phi'(\varepsilon)
&=&
-\nu\frac{(2-a)(1-2\lambda(2-a))}{g({\varepsilon})}\phi^{(0)}
+{\cal O}(\nu^2, k^2)
\nonumber
\\
&=&
i\omega\bigg(\frac{l^2}{2Nr_+}\bigg)
\frac{1-2\lambda(2-a)}{g(\varepsilon)}\phi^{(0)}
+{\cal O}(\omega^2, k^2).
\end{eqnarray}
Therefore we can read off the correlation function from the
relation (\ref{green}),
\begin{equation}
G_{xy \ xy}(\omega, k)
=-i\omega\frac{1}{16\pi G_5}\left(\frac{r_+^3}{l^3}\right)
\Big(1-2\lambda(2-a)\Big)
+{\cal O}(\omega^2, k^2),
\end{equation}
where we subtracted contact terms.
Then finally, we can obtain the shear viscosity by using Kubo formula
(\ref{kubo}),
\begin{equation}
\eta=\frac{1}{16 \pi
G_5}\left(\frac{r^3_{+}}{l^3}\right)\Big(1-2\lambda (2-a)\Big).
\end{equation}
The ration of the shear viscosity to the entropy density
is concluded as
\begin{equation}
\frac{\eta}{s}
=\frac{1}{4 \pi }\left(1-4\lambda (1-\frac{a}{2})\right).
\end{equation}
One can see explicitly that the conjectured viscosity bound can be
violated for some value of $\lambda$ and $a$. Figure \ref{viscosity}
demonstrates that for fixed value of $\eta/s$, as the couping
constant $\lambda$ increases, $a$ also increases. The shear
viscosity approaches zero as $(\lambda,a)\rightarrow (1/4,0)$ and
$\lambda$ is thus bounded by $1/4$. When $a=0$ (no charges),
$\eta/s=(1-4\lambda)/(4\pi)$, we recover the result in
Ref.\cite{shenker}. It is also worth noting that for extremal case
($a=2$), the ratio of the shear viscosity to entropy density
receives no corrections from Gauss-Bonnet terms.

\section{Causality violation}
\setcounter{equation}{0}
\setcounter{footnote}{0}

It was shown that the causality could be violated if one introduced
Gauss-Bonnet terms~\cite{shenker, shenker1}.
We here analyze an effect of the charge to this issue.

Due to higher derivative terms in the gravity action,
the equation (\ref{maineq})
for the propagation of a transverse
graviton differs from that of a minimally coupled
massless scalar field
propagating in the same background geometry.
Writing the wave function as
\begin{equation}
\label{phi}
\phi(x,u)=\mbox{e}^{-i\omega t+ikz+ik_{u}u},
\end{equation}
and taking large momenta limit $k^\mu\rightarrow\infty$, one can
find that the equation of motion (\ref{maineq}) reduces to
\begin{equation}
\label{effeq}
k^{\mu}k^{\nu}g_{\mu\nu}^{\rm eff}\simeq 0,
\end{equation}
where the effective metric is given by
\begin{equation}
\dd s^2_{\rm eff}
=g^{\rm eff}_{\mu\nu}\dd x^\mu\dd x^\nu
=\frac{N^2 f(u)}{l^2 b^2 u}
\left(-\dd t^2+\frac{1}{c^2_g}\dd z^2\right)
+\frac{l^2}{4 u^2 f(u)}\dd u^2.
\end{equation}
Note that $c^2_g$ can be interpreted as the local speed of graviton:
\begin{equation}
c^2_g(u)=\frac{N^2 f(u)
\bigg[1-2\lambda u^2\Big(2u(u^{-1}f(u))''+3(u^{-1}f(u))'\Big)\bigg]}
{1+2\lambda u^2(u^{-1}f(u))'}.
\end{equation}
We can expand $c^2_g$ near the boundary $u=0$,
\begin{eqnarray}
c^2_g-1=
\left(-\frac{5}{2}(1+a)+\frac{2(1+a)}{1-4\lambda}
-\frac{1+a}{2\sqrt{1-4\lambda}}\right)u^2
+\mathcal{O}(u^3).
\end{eqnarray}
%
%As the local speed of graviton should be smaller than $1$
% (the local speed of light), we require
%
As we will see below, the local speed of graviton should be smaller than $1$
(the local speed of light of boundary CFT).
We require
\begin{equation}
-\frac{5}{2}+\frac{2}{1-4\lambda}
-\frac{1}{2\sqrt{1-4\lambda}}\leq 0.
\end{equation}
The above equation leads to $\lambda \leq 0.09$
without any charge dependence.
This is the same result with neutral black holes in
Gauss-Bonnet theory\cite{shenker1}.
For charged black branes, the speed of graviton
$c^2_g$ is also smaller than $1$ in the range $\lambda \leq 0.09$.

\begin{figure}[htbp]
\begin{minipage}{1\hsize}
\begin{center}
\includegraphics*[scale=0.48]{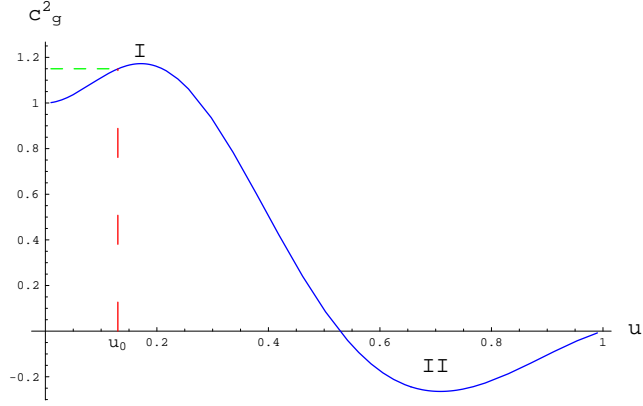}
\end{center}
\caption{The hump (I) signifies the causality violation,
while the well (II) indicates that the black brane is unstable.}
\label{geodesic}
\end{minipage}
\end{figure}

Now let us study in the regime $\lambda > 0.09$, to see how the
causality is violated in the boundary theory. We follow the
discussion in the papers~\cite{shenker, shenker1}. By using the
identification $\displaystyle\frac{\dd x^{\mu}}{\dd s} =g^{{\rm
eff}\mu\nu}k_{\nu}$, we may rewrite the equation (\ref{effeq}) into
one for geodesic motion,
\begin{equation}
\label{geod}
\frac{\dd x^{\mu}}{\dd s}\frac{\dd x^{\nu}}{\dd s}g^{\rm eff}_{\mu\nu}=0.
\end{equation}
If we consider $\omega$ and $q$
as conserved integrals of motion along the geodesic with
\begin{equation}
\omega=\left(\frac{\dd t}{\dd s}\right)\frac{f N^2}{l^2b^2u},
\qquad
k=\left(\frac{\dd z}{\dd s}\right)\frac{f N^2}{l^2b^2u}\frac{1}{c^2_g},
\end{equation}
from (\ref{effeq}) and (\ref{geod}), we can obtain
\begin{equation}
\left(\frac{N}{k^2 b}\frac{\dd u^{-\frac{1}{2}}}{\dd s}\right)^2
=\frac{\omega^2}{k^2}-c^2_g.
\end{equation}
We can simplify the above equation by recalling $s$ as $\tilde{s}=ks/N$
and noting that $u=1/({r^2 b^2})$,
\begin{equation}
\label{m}
\left(\frac{\dd r}{\dd\tilde{s}}\right)^2=\alpha^2-c^2_g,
\qquad
\alpha^2=\frac{\omega^2}{k^2}.
\end{equation}
The geodesic equation determines the radial motion of a test
particle with energy $\alpha^2$ in an effective potential
$c^2_g$.
One can now infer that, from Figure \ref{geodesic},
geodesic line which starts at spatial
infinity (the boundary) can bounce back to the boundary.
In other words, the turning point appears at
\begin{equation}
\alpha^2=c^2_g (u_0).
\end{equation}

\noindent
For a light-like geodesic line starting from the boundary
and bounced back at the boundary, we find
\begin{subequations}
\begin{eqnarray}
\label{t}
\Delta t
&=&
2\!\int^{\infty}_{r_0}
\frac{\dd t}{\dd\tilde{s}}\frac{\dd\tilde{s}}{\dd r}\dd r
=\frac{2}{N}\!\int^{\infty}_{r_0}
\!\dd r \frac{\alpha}{f\sqrt{\alpha^2-c^2_g}},
\\
\label{x3}
\Delta z
&=&
2\!\int^{\infty}_{r_0}
\frac{\dd z}{\dd\tilde{s}}\frac{\dd\tilde{s}}{\dd r}\dd r
=\frac{2}{N}\!\int^{\infty}_{r_0}
\!\dd r \frac{c^2_g}{f\sqrt{\alpha^2-c^2_g}}.
\end{eqnarray}
\end{subequations}

\vspace*{-6mm}
\noindent
As pointed out in Ref.\cite{shenker1}, one may find microcausality
violation in the boundary  CFT when the a bouncing graviton geodesic
satisfies $\displaystyle\frac{\Delta z}{\Delta t}> 1$.
From (\ref{t}) and (\ref{x3}), we can see that if we tune
$c_g(u_0)$ to be $c_{g,{\rm max}}$, we then have
\begin{equation}
\frac{\Delta z}{\Delta t}\rightarrow c_{g, {\rm max}}> 1.
\end{equation}
Since near the boundary $c_g$ can be greater than $1$, the
propagation of signals in the boundary theory with speed
$\displaystyle\frac{\Delta z}{\Delta t}$ might become superluminal.
Actually, metastable states can live in  the well between $c^2_{g,
{\rm max}}$ and the boundary and $\displaystyle\frac{\Delta
z}{\Delta t}$ corresponds to the group velocity of metastable
quasiparticles. According to the standard procedure to analyze
quasinormal modes, we rewrite the wave function in a Schr\"odinger
form,
\begin{equation}
\label{sch}
-\frac{\dd^2 \psi}{\dd r^2_{*}}+V\left(r(r_{*})\right)\psi
=\bar{\omega}^2 \psi,
\qquad
\frac{\dd r_{*}}{\dd r}=\frac{1}{Nl^2b^2 f(r)},
\end{equation}
where $\psi(r(r_*))$ and the potential is defined by
\begin{eqnarray*}
\psi
&=&
K(r)\phi,
\qquad
K(r)\equiv\sqrt{\frac{g(u)}{u^{-1}f(u)}}
=1-\lambda br\frac{\partial(l^2b^2 f(r))}{\partial r},
\\
V
&=&
k^2c^{2}_g+V_{1}(r),
\qquad
V_{1}(r)\equiv{N^2}l^2b^2 \left[\left(f(r)\frac{\partial \ln K(r)}
{\partial r}
\right)^2+f(r)\frac{\partial}{\partial r}
\left(f(r)\frac{\partial \ln K(r)}{\partial r}\right)\right].
\end{eqnarray*}
Following the procedure of Ref.\cite{shenker1},
one can find that the group velocity of the graviton is given by
\begin{equation}
v_g=\frac{\dd\omega}{\dd k}=\frac{\Delta z}{\Delta t}.
\end{equation}
Therefore, signals in the boundary theory propagate outside of the light cone.
Again, we confirm that microcausality violation happens in the CFT.
One may expect that when $\lambda\leq 0.09$, the theory with
Gauss-Bonnet corrections is safe and consistent.

\section{Instability}
\setcounter{equation}{0}
\setcounter{footnote}{0}

Apart from the causality violation, for RN-AdS black brane in Gauss-Bonnet
theory, the charges give  the
instability of the black brane within the window of
$0<\lambda\leq 0.09$.

\begin{figure}[htbp]
\begin{minipage}{1\hsize}
\begin{center}
%\vspace*{10mm}
\includegraphics*[scale=0.53] {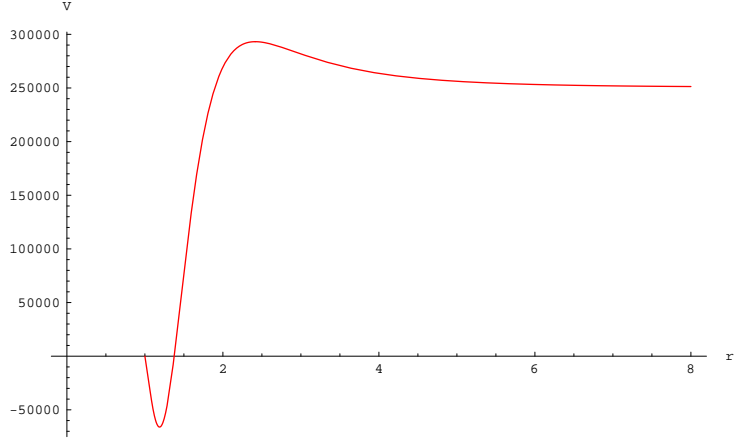}
\end{center}
\caption{Schr\"odinger potential $V(r)$ as a function of $r$
for $\lambda=0.2$, $a=1.7$ and $k=500$.
Near the horizon, the potential develops a negative-valued well and
then negative-energy bound states will appear. Near the boundary,
the potential develops a hump which corresponds to superluminal
propagation of metastable quasiparticles.}
\label{potential}
\end{minipage}
\end{figure}

From Figure \ref{potential}, we can see that the Schr\"odinger
potential develops a negative gap near the horizon.
We will now show that in the large momentum limit, the
negative-valued potential leads to instability of the black brane.
In the large momenta limit $k^\mu\rightarrow \infty$, the dominant
contribution to the potential is given by $k^2 c^2_g$. For near
extremal cases, $c^2_g$ can be negative near the horizon and
$V\simeq k^2 c^2_g$ can be deep enough (see Figure \ref{potential}).
Thus bound states can live in the negative-valued well. The negative
energy bound state corresponds to modes of tachyonic mass on
Minkowski slices \cite{troost} and signals an instability of the
black brane~\cite{dotti,konoplya}. Let us expand $c^2_g$ in series
of $(1-u)$,
\begin{equation}
c^2_g=\frac{(2-a)
\Big(1+4 \lambda -14 a \lambda-32 \lambda^2+32a\lambda^2-8a^2 \lambda^2\Big)}
{(1-4\lambda+a \lambda)}\big(1-u\big)+\mathcal{O}\left((1-u)^2\right).
\end{equation}
Since $0\leq a\leq 2$, and $0\leq u \leq 1$, $c^2_g$ will be negative,
if
\begin{equation}
\frac{ (1+4 \lambda -14 a \lambda-32 \lambda^2+32a\lambda^2-8a^2 \lambda^2)}
{(1-4\lambda+2 a \lambda)} <0.
\end{equation}
From the above formula, we find the critical value of $\lambda$,
\begin{equation}
\lambda_{\rm c} = \frac{2-7a+\sqrt{3}\sqrt{12-20 a+19 a^2}}{8(a-2)^2}.
\end{equation}
Above the line of $\lambda_{\rm c}$, $c^2_g$ can be negative
(see figure 2).
The minimal value of $\lambda_{\rm c}$ can be obtained in the
limit $a\rightarrow 2$,
\begin{equation}
\lambda_{\rm c, \ min} = \frac{1}{24}.
\end{equation}

\begin{figure}[bpht]
\begin{minipage}{1\hsize}
\vspace*{5mm}
\begin{center}
\includegraphics*[scale=0.45]{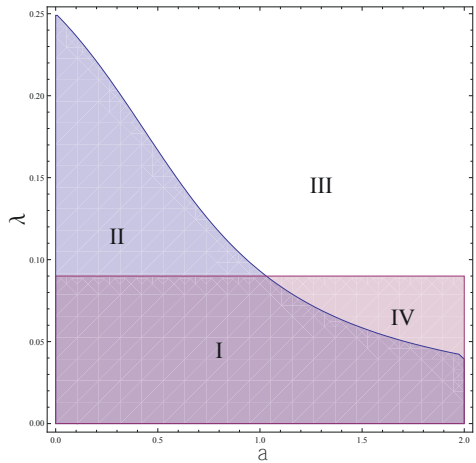}
\end{center}
\caption{ Phase diagram for the instability in $a$-$\lambda$ space.
Region I: There are no causality violation and bulk bound states.
Region II: Causality violation can happen, but bound states do not
appear in the bulk. Region III: Both causality violation and
instability happen in this region. Region IV: There is no causality
violation, but the black brane is unstable.} \label{phase}
\end{minipage}
\end{figure}

\begin{figure}[bpht]
\begin{minipage}{0.47\hsize}
\vspace*{5mm}
\begin{center}
\includegraphics*[scale=0.47]{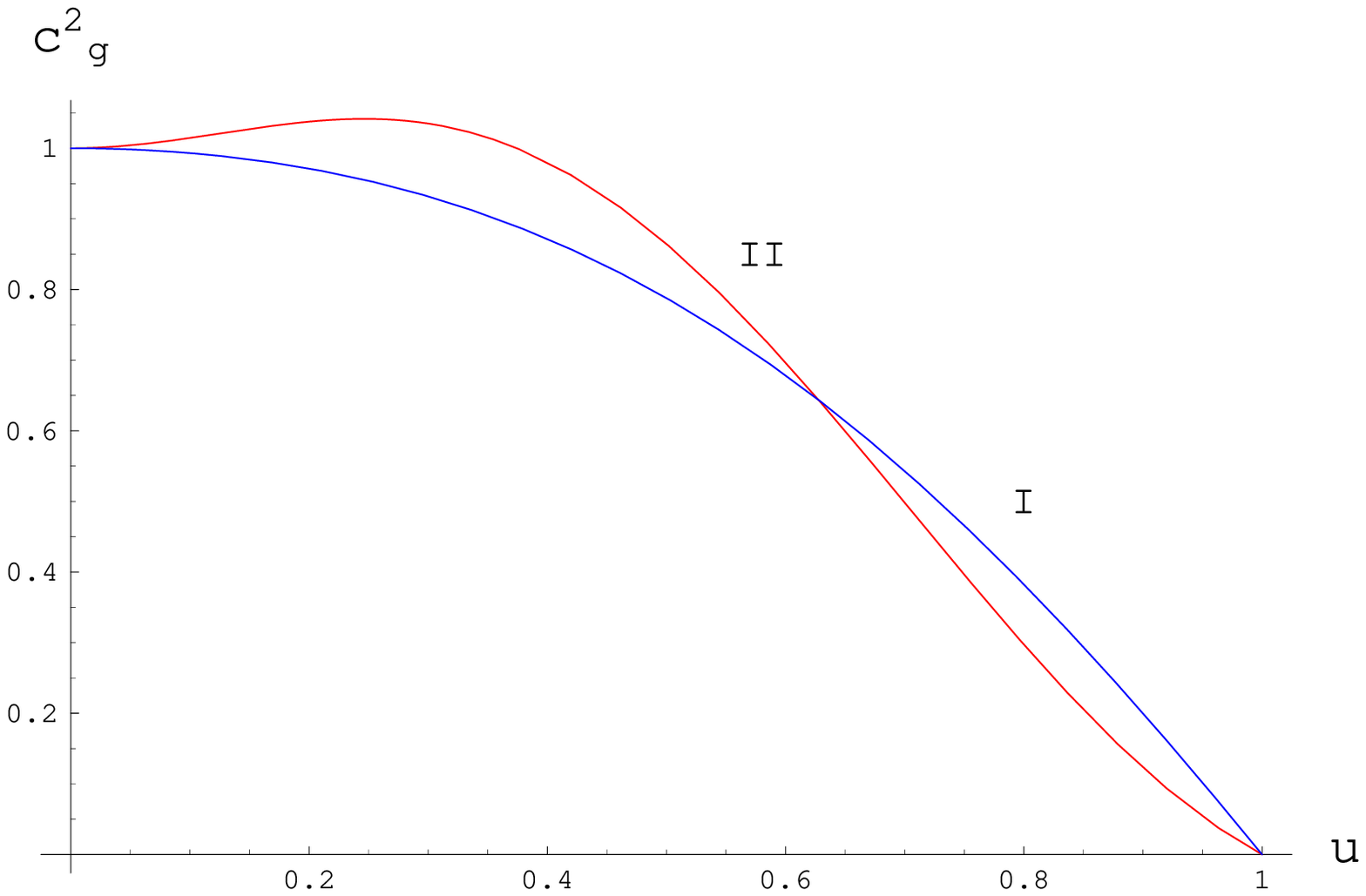}
\end{center}
\caption{$c^2_g$ as a function of $u$. Line I  describes the
behavior of $c^2_g$ in region I for $\lambda=0.05$, $a=0.2$. Line II
shows that  $c^2_g>1$ at some value of $u$ near the boundary for
$\lambda=0.14$, $a=0.4$.} \label{cg1}
\end{minipage}
\quad
\begin{minipage}{0.47\hsize}
\vspace*{10mm}
\begin{center}
\includegraphics*[scale=0.47]{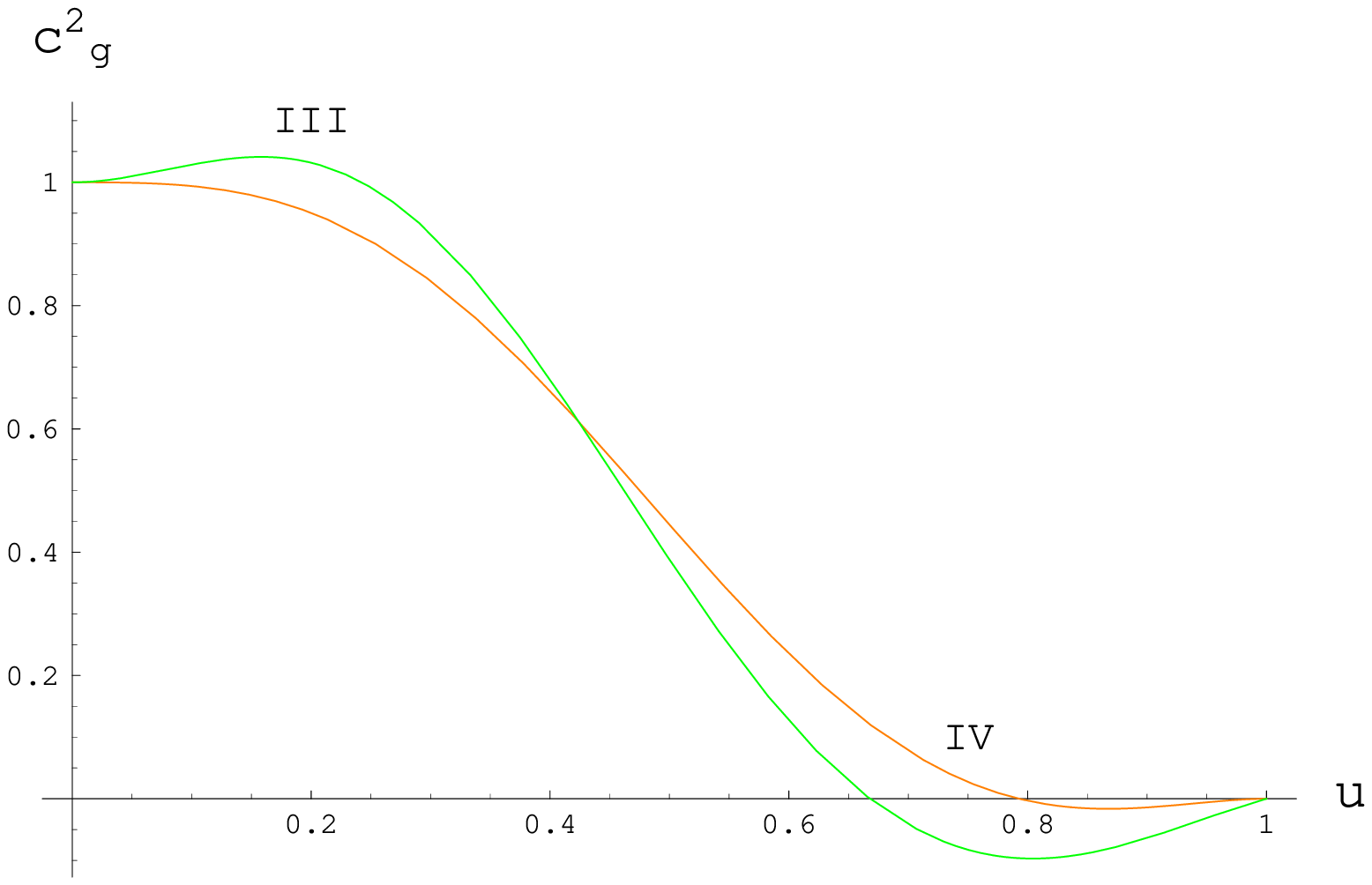}
\end{center}
\vspace*{-2mm}
\caption{$c^2_g$ as a function of $u$. Line III: $c^2_g$ has a hump
greater than 1 and a negative-valued well (We set $\lambda=0.15$,
$a=1.7$).
 Line IV: $c^2_g$ only has a negative-valued well ($\lambda=0.089$, $a=1.99$) }
 \label{cg2}
\end{minipage}
 \end{figure}

Figure \ref{phase} shows us that the two lines
$\lambda_{\rm c}(a)$ and $\lambda=0.09$
separates the physics into four regions in $(a,\lambda)$ space.
The physics in region I so far can be consistent.
In region II, causality violation can be found.
In region III, causality violation
as well as unstable quasinormal modes (QNMs) appear.
In region IV, we can only find unstable QNMs.
Figures \ref{cg1} and \ref{cg2} show us explicitly
the behaviors of $c^2_g$ in different regions.
In order to demonstrate that the peculiar feature of $c^2_g< 0$
signals the instability of the RN-AdS black brane in Gauss-Bonnet theory,
we solve the Schr\"odinger equation (\ref{sch})
with negative-valued potential numerically and find some unstable
QNMs (see Table 1).
\begin{table*}[htbp]
\caption{Unstable QNMs for charged GB black brane perturbation of
tensor type.}
\begin{center}
\begin{tabular}{|c|c|c|c|c|c|c|c|c|}
\hline
$\lambda$&$a=1.9$&$a=1.7$&$a=1.4$&$a=1.1$&$a=0.8$&$a=0.5$\\
\hline
$0.2$&$ 258.2823\imo$&$255.9158 \imo$&$  245.6454\imo$&$223.4933\imo$&$179.9353\imo$&$94.4075\imo$\\
$0.15$&$ 158.0242 \imo$&$154.8652\imo$&$ 141.6929\imo$&$ 113.4577\imo$&$ 60.5891\imo$&$ \rm - $\\
$0.1$&$ 79.3897 \imo$&$ 75.4296\imo$&$ 59.4755\imo$&$ 26.9101\imo$&$ -$&$ \rm -$\\
$0.07$&$33.9841 \imo$&$ 29.3078\imo$&$- $&$ -$&$ -$&$ -$\\
$0.05$&$ 6.6761\imo$&$ -$&$ -$&$ -$&$ -$&$ -$\\
\hline
\end{tabular}
\end{center}
\end{table*}
From Table 1, we can find that the real part of $\omega$ is vanishing,
while the imaginary part of  $\omega$  is positive.
Inserting the values of $\omega$ into the equation (\ref{phi}),
one can see that gravitational instability grows
as time goes on and then the black brane
becomes unstable against gravitational perturbation.
The numerical analysis also indicate that the black brane
becomes stable under gravitational perturbation,
when we restrict $\lambda$ to be $\lambda\leq {1}/{24}$.
%
%That is to say, the stability of the charged black brane for
%all charge requires $\eta/s\geq (5/6)(1/4\pi)$.

\section{Conclusions and discussions}
\setcounter{equation}{0}
\setcounter{footnote}{0}

In summary,
we have computed the charge dependence of ${\eta}/{s}$ for
Gauss-Bonnet theory.
We have taken RN-AdS black brane solution
into Gauss-Bonnet gravity
and used the Kubo formula to compute the viscosity of the dual boundary
theory.
The ratio of the shear viscosity to
entropy density was found to be
${\eta}/{s}=
\Big(1-4\lambda(1-a/2)\Big)/(4\pi)$,
which violated the conjectured
viscosity bound for non-extremal RN-AdS black brane in Gauss-Bonnet gravity.
However, for extremal case, the conjectured
lower viscosity-entropy density bound $1/4\pi$ can be recovered.

The causality violation and the instability of charged black brane
were also analyzed in this paper. We have confirmed the results
found in previous work that when $\lambda > 0.09$, causality
violation happens in the boundary CFT \cite{shenker1}. It is
interesting to notice that charges introduce instability of the
black brane even in the range $0<\lambda\leq 0.09$. Therefore, to
avoid causality violation and instability for any charge, we suggest
to restrict the value of $\lambda$ to be
 $\lambda\leq {1}/{24}\sim0.04$.
%Consequently, the ratio might be $\eta/s\ge (5/6)(1/4\pi)$.

Our final remark is about the interpretation of the charge effect.
If one introduces the bulk filling D-branes, one can consider the
two kind of Maxwell fields, one for $R$-charge $U(1)$ and the other
for the baryon charge for the brane gauge field~\cite{sinRN}. Since
both fields couple to the gravity in the same way, one can consider
the charge in our analysis either as the $R$-charge or baryon
charge. For the latter case, we can interpret the charge effect as
the effect of finite baryon density.

\vspace*{10mm}
\noindent
 {\large{\bf Acknowledgments}}

\vspace{1mm}
XHG, YM, FWS and TT would like to thank CQUeST and Hanyang University
for warm hospitality.
The work of SJS was supported by KOSEF Grant R01-2007-000-10214-0.
This work is also supported by Korea Research Foundation Grant
KRF-2007-314-C00052 and SRC Program of the KOSEF
through the CQUeST with grant number R11-2005-021.


\begin{thebibliography}{99}
\bibitem{ads/cft}
J.M. Maldacena,
{Adv. Theor. Math. Phys.} {\bf 2} (1998) 231,
{\tt [arXiv:hep-th/9711200]}.
\bibitem{gkp}
S.S. Gubser, I.R. Klebanov and A.M. Polyakov,
Phys.\ Lett.\ {\bf B428} (1998) 105,
{\tt [arXiv:hep-th/9802109]}.
\bibitem{w}
E. Witten,
Adv.\ Theor.\ Math.\ Phys.\ {\bf 2} (1998) 253,
{\tt [arXiv:hep-th/9802150]}.
\bibitem{pss0}
G. Policastro, D.T. Son and A.O. Starinets,
Phys.\ Rev.\ Lett.\  {\bf 87} (2001) 081601,
{\tt [arXiv:hep-th/0104066]}.
\bibitem{kss}
P. Kovtun, D.T. Son and A.O. Starinets,
JHEP {\bf 0310} (2003) 064, \\
{\tt [arXiv:hep-th/0309213]}.
\bibitem{bl}
A. Buchel and J.T. Liu,
Phys.\ Rev.\ Lett.\  {\bf 93} (2004) 090602, \\
{\tt [arXiv:hep-th/0311175]}.
\bibitem{ssz}
E. Shuryak, S.-J. Sin and I. Zahed,
J.\ Korean Phys.\ Soc.\  {\bf 50} (2007) 384, \\
{\tt [arXiv:hep-th/0511199]}.
\bibitem{ksz}
K.-Y. Kim, S.-J. Sin and I. Zahed,
{\tt [arXiv:hep-th/0608046]}.
\bibitem{ht}
N. Horigome and Y. Tanii,
JHEP {\bf 0701} (2007) 072,
{\tt [arXiv:hep-th/0608198]}.
\bibitem{nssy1}
S. Nakamura, Y. Seo, S.-J. Sin and K.P. Yogendran,
{\tt [arXiv:hep-th/0611021]}.
\bibitem{kmmmt}
S. Kobayashi, D. Mateos, S. Matsuura, R.C. Myers and R.M. Thomson,
JHEP {\bf 0702} (2007) 016,
{\tt [arXiv:hep-th/0611099]}.
\bibitem{nssy2}
S. Nakamura, Y. Seo, S.-J. Sin and K.P. Yogendran,
{\tt [arXiv:0708.2818[hep-th]]}.
\bibitem{kovtun}
P. Kovtun, D.T. Son and A.O. Starinets,
Phys.\ Rev.\ Lett.\  {\bf 94} (2005) 111601,
{\tt [arXiv:hep-th/0405231]}.
\bibitem{kp}
Y. Kats and P. Petrov,
{\tt [arXiv:0712.0743[hep-th]]}.
\bibitem{shenker}
M. Brigante, H. Liu, R.C. Myers, S. Shenker and S. Yaida,
Phys. Rev. {\bf D77} (2008) 126006, {\tt [arXiv:0712.0805[hep-th]]}.
\bibitem{shenker1}
M. Brigante, H. Liu, R.C. Myers, S. Shenker and S. Yaida,
Phys. Rev. Lett. {\bf 100} (2008) 191601, {\tt [arXiv:0802.3318[hep-th]]}.
\bibitem{neupane}
I.P. Neupane and N. Dadhich, {\tt [arXiv:0808.1919[hep-th]]}.
\bibitem{dotti}
G. Dotti and R.J. Gleiser,
Phys.\ Rev.\ {\bf D72} (2005) 044018,
{\tt [arXiv:gr-qc/0503117]}; \\
R.J. Gleiser and G. Dotti,
Phys.\ Rev.\  {\bf D72} (2005) 124002,
{\tt [arXiv:gr-qc/0510069]}; \\
M. Beroiz, G. Dotti and  R.J. Gleiser,
Phys. Rev. {\bf D76} (2007) 024012, \\
{\tt [arXiv:hep-th/0703074]}.
\bibitem{konoplya}
R.A. Konoplya and A. Zhidenko,
Phys. Rev. {\bf D77} (2008) 104004,
{\tt [arXiv:0802.0267]}.
\bibitem{g1}
D. G. Boulware and S. Deser,
Phys. Rev. Lett. {\bf 55} (1985) 2625.
\bibitem{g2}
D. Whiltshire,
Phys. Rev. {\bf D38} (1988) 2445.
\bibitem{g3}
R.G. Cai,
Phys. Rev. {\bf D65} (2002) 084014, {\tt [arXiv:hep-th/0109133]};
\\
R.G. Cai and Q. Guo,
Phys. Rev. {\bf D69} (2004) 104025, {\tt [arXiv:hep-th/0311020]}.
\bibitem{cvetic}
M. Cveti\v{c}, S. Nojiri and S.D. Odintsov,
Nucl. Phys. {\bf B628} (2002) 295, \\
{\tt [arXiv:hep-th/0112045]}.
\bibitem{g4}
I.P. Neupane, Phys. Rev. {\bf D67} (2003) 061501,
{\tt [arXiv:hep-th/0212092]};
Phys. Rev. {\bf D69} (2004) 084011,
{\tt [arxiv:hep-th/0302132]}.
\bibitem{ast}
D. Astefanesei, N. Banerjee  and S. Dutta,
 {\tt  [arXiv:0806.1334 [hep-th]]}.
\bibitem{gmsst}
X.-H. Ge, Y. Matsuo, F.-W. Shu, S.-J. Sin and T. Tsukioka, \\
{\tt [arXiv:0806.4460[hep-th]]}.
\bibitem{w2}
E. Witten,
Adv.\ Theor.\ Math.\ Phys.\ {\bf 2} (1998) 505,
{\tt [arXiv:hep-th/9803131]}.
\bibitem{ss}
D.T. Son and A.O. Starinets, JHEP {\bf 0209} (2002) 043,
{\tt [arXiv:hep-th/0205051]}.
\bibitem{troost}
J. Troost,
Phys. Lett. {\bf B578} (2004) 210, {\tt [arXiv:hep-th/0308044]};
\\
P. Minces,
Phys. Rev. {\bf D70} (2004) 025011, {\tt [arXiv:hep-th/0402161]}.
\bibitem{sinRN}
S.-J. Sin, JHEP {\bf 0710} (2007) 078,
{\tt [arXiv:0707.2719[hep-th]]}.

\end{thebibliography}
\end{document}